\documentclass[conference]{IEEEtran}

\usepackage{fst}
\usepackage{cite}

\begin{document}

\title{Optimization of Bit Mapping and Quantized Decoding for Off-the-Shelf Protograph LDPC Codes with Application to IEEE 802.3ca}

\author{\IEEEauthorblockN{Fabian Steiner, Gerhard Kramer}
\IEEEauthorblockA{Institute for Communications Engineering\\Technical University of Munich\\Email: \{fabian.steiner, gerhard.kramer\}@tum.de}
}

\maketitle

\begin{abstract}
Protograph-based, off-the-shelf low-density parity-check (LDPC)
codes are optimized for higher-order modulation and quantized sum-product decoders.
As an example, for the recently proposed LDPC code from the upcoming IEEE 802.3ca standard for
passive optical networks (PONs), an optimized mapping of the bit channels originating from bit-metric decoding 
to the protograph variable nodes gains \SI{0.4}{dB} and \SI{0.3}{dB} at a bit-error rate of
\num{e-6} for shaped and uniform signaling, respectively. Furthermore, the clipping value for a quantized
sum-product LDPC decoder is optimized via discretized density evolution.
\end{abstract}

\section{Introduction}

Passive optical networks (PONs)\acused{PON} use unpowered fiber optic splitters to 
serve multiple terminals with one common optical fiber. Their predominant use is in \ac{FTTH} networks. Most PONs operate in a \ac{TDMA} manner
and provide up to \SI{10}{Gbit/s} with simple noncoherent \ac{OOK} modulation.
New standards for \SI{25}{Gbit/s}, \SI{50}{Gbit/s} and \SI{100}{Gbit/s} \acp{PON} are underway, which 
will result in the IEEE 802.3ca standard in 2019. The standardization consortium agreed on \ac{LDPC} codes 
\cite{ieee802.3ca} as the preferred \ac{FEC} solution.

Coherent detection and higher-order modulation formats such as $M$-\ac{QAM} are becoming increasingly important also
for \ac{PON} networks to enable a greater flexibility in data rates~\cite{van_der_linden_adaptive_2018}.
To operate \ac{LDPC} codes with higher order modulation, usually \ac{BMD} is employed, where the demapper calculates a bit-wise soft information 
for each of the $m = \log_2(M)$ bits indexing a constellation symbol. \ac{BMD} is also the key component of \ac{BICM}~\cite{guillen_i_fabregas_bit-interleaved_2008}.
Various optimization techniques have been proposed to map the \ac{BMD} bit levels to the different \ac{VN} degrees of
an irregular \ac{LDPC} code or to the \ac{VN} types of a protograph~\cite{thorpe_protograph}. Two approaches can be distinguished in literature. The first approach
considers the bit mapping optimization for an off-the-shelf \ac{LDPC} code~\cite{divsalar_protograph_2005-3,richter_mapping_2007,jin_optimized_2010,cheng_exit-aided_2012,hager2014optimized,hager_improving_2014}. 
The second designs the code and the bit mapping jointly~\cite{zhang_MET_journal,steiner_protograph-based_2016}. While the second approach can
exploit all degrees of freedom, practical constraints often require to reuse a given code for a new scenario and the best bit mapping
needs to be found. This is the case when the proposed \ac{PON} \ac{LDPC} code is used for higher order modulation.

\begin{figure}[t]
 \footnotesize
 \centering
 \begin{tikzpicture}
\begin{axis}[
xlabel={$\text{SNR}~[\si{dB}]$},
ylabel={BER},
grid=both,
ymode=log,
legend pos=outer north east,
ymin=1e-9,
max space between ticks=20
]

\addplot[name path global=shaped,line width=1,TUMBeamerRed,mark=o,dashed,mark options={solid}] table[x=snr,y=ber] {data/results-pon-shaped-consecutive_R=2.55_I=100.txt};\label{plt:shaped_consecutive}

\addplot[name path global=shaped_opt,line width=1,TUMBeamerRed,mark=o] table[x=snr,y=ber] {data/results-pon-shaped_R=2.55_I=100.txt};\label{plt:shaped_opt}

\addplot[name path global=uni,line width=1,TUMBeamerBlue,mark=o,dashed,mark options={solid}] table[x=snr,y=ber] {data/results-pon-consecutive_R=2.55_I=100.txt};\label{plt:uni_consecutive}
\addplot[name path global=uni_opt,line width=1,TUMBeamerBlue,mark=o] table[x=snr,y=ber] {data/results-pon-R=2.55_I=100.txt};\label{plt:uni_opt}

\addplot[name path global=shaped_7,line width=1,TUMBeamerRed,mark=diamond] table[x=snr,y=ber] {data/results-quant-shaped-B=6.00_q=7.txt};\label{plt:shaped_7}

\addplot[name path global=shaped_15,line width=1,TUMBeamerRed,mark=square] table[x=snr,y=ber] {data/results-quant-shaped-B=8.00_q=15.txt};\label{plt:shaped_15}

\addplot[name path global=shaped_15,line width=1,TUMBeamerRed,mark=+] table[x=snr,y=ber] {data/results-quant-shaped-B=15.00_q=255.txt};\label{plt:shaped_255}

\draw[gray] (16.4827,1e-9) -- (16.4827,1e-3) node[near start,sloped,above,xshift=0.35cm] {\small BMD limit uniform};

\draw[gray] (15.25,1e-9) -- (15.25,1e-3) node[near start,sloped,above,xshift=0.1cm] {\small BMD limit PAS};

\path[name path global=line] (10,1e-6) -- (20,1e-6);

\path[name intersections={of=line and uni, name=p1}, name intersections={of=line and uni_opt, name=p2}];

\draw[<->,thick] let \p1=(p2-1), \p2=(p1-1) in (p1-1) -- (p2-1) node [below,midway,fill=white,yshift=0.5cm,xshift=.7cm] {%
        \pgfplotsconvertunittocoordinate{x}{\x1}%
        \pgfplotscoordmath{x}{datascaletrafo inverse to fixed}{\pgfmathresult}%
        \edef\valueA{\pgfmathresult}%
        \pgfplotsconvertunittocoordinate{x}{\x2}%
        \pgfplotscoordmath{x}{datascaletrafo inverse to fixed}{\pgfmathresult}%
        \pgfmathparse{\pgfmathresult - \valueA}%
        \pgfmathprintnumber{\pgfmathresult} dB
};

\path[name intersections={of=line and shaped, name=p1}, name intersections={of=line and shaped_opt, name=p2}];

\draw[<->,thick] let \p1=(p2-1), \p2=(p1-1) in (p1-1) -- (p2-1) node [below,midway,fill=white,yshift=0.5cm,xshift=1cm] {%
        \pgfplotsconvertunittocoordinate{x}{\x1}%
        \pgfplotscoordmath{x}{datascaletrafo inverse to fixed}{\pgfmathresult}%
        \edef\valueA{\pgfmathresult}%
        \pgfplotsconvertunittocoordinate{x}{\x2}%
        \pgfplotscoordmath{x}{datascaletrafo inverse to fixed}{\pgfmathresult}%
        \pgfmathparse{\pgfmathresult - \valueA}%
        \pgfmathprintnumber{\pgfmathresult} dB
};

\end{axis}
\end{tikzpicture}
 \caption{Comparison of uniform (reference: \ref{plt:uni_consecutive}, optimized: \ref{plt:uni_opt} bit mapping) and PAS signaling (reference: \ref{plt:shaped_consecutive}, optimized: \ref{plt:shaped_opt} bit mapping) for 
 a target spectral efficiency of $\eta = \SI{2.545}{bpcu}$. Uniform signaling uses an 8-ASK constellation, while PAS uses 16-ASK. For the optimized bit mapping with PAS, we also show the performance of
 a quantized decoder with three (\ref{plt:shaped_7}), four (\ref{plt:shaped_15}) and eight bits (\ref{plt:shaped_255}).}
 \label{fig:cmp_uni_shaped}
\end{figure}
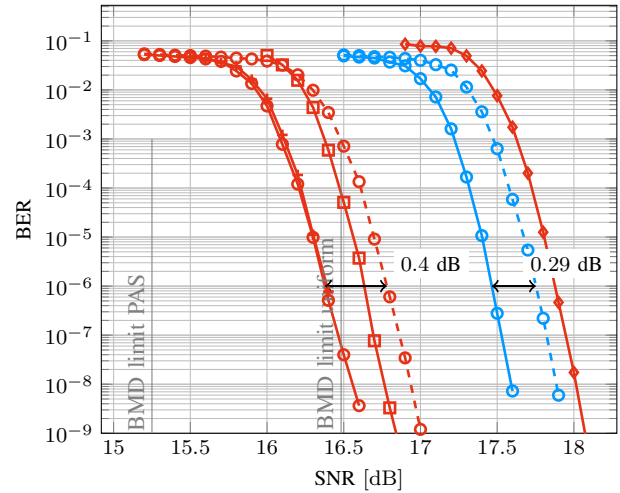

None of the previously suggested optimization approaches is tailored to the code defined in IEEE 802.3ca and also other standards (e.g., IEEE 802.11,
G.hn): Some assume \emph{unstructured} LDPC codes (e.g., \cite{richter_mapping_2007,cheng_exit-aided_2012}) and others are prohibitively complex to work 
with the protograph sizes in standards. For example, the authors of~\cite{hager2014optimized} use differential evolution to optimize the bit mapping for
protograph based spatially coupled \ac{LDPC} codes with window decoding and exploit the periodicity imposed by window decoding to limit the optimization space. Furthermore,
sampling from the high-dimensional polytope to generate populations for differential evolution becomes rather time consuming.

In this work, we propose an algorithm that optimizes the bit mapping of protograph based \ac{LDPC} codes one level after the other.
We use the surrogate approach of \cite{steiner_protograph-based_2016}  and P-EXIT~\cite{liva_protograph_2007} analysis
to determine the decoding threshold for a given mapping and use the \emph{patternsearch} algorithm~\cite{hooke_direct_search_1961}
to find the best bit mapping. We validate this approach by comparing the predicted P-EXIT thresholds with \ac{DDE}. 

Additionally, practical \ac{LDPC} decoders often quantize the exchanged messages with a finite number of bits. This is particularly important
for optical communication with its high throughput requirements~\cite[Sec.~III]{smith_staircase_2012}.
Previous works~\cite{chen_density_2002,zhao_implementation_2005} noted that the performance of quantized decoders depends on the
clipping of messages and found the optimal clipping by time consuming finite length simulations. 

As a second contribution, we therefore optimize the clipping of a quantized sum-product \ac{LDPC} decoder exemplarily for three and four bits resolution.
We use the decoding threshold of an ensemble as the objective and show that \ac{DDE} accurately predicts the finite length performance,
making it an important tool to faciliate the design process.

The paper is structured as follows. In Sec.~\ref{sec:prelim} we present the system model, introduce the basic information theoretic quantities
and provide a brief introduction to \ac{LDPC} codes and their asymptotic analysis with \ac{DDE}. Sec.~\ref{sec:optim_mapping} formalizes the bit mapping optimization problem and presents our successive 
bit allocation mapping approach. In Sec.~\ref{sec:opt_clipping}, we illustrate the effect of clipping for quantized \ac{LDPC} decoders and optimize it for different number of bits. We conclude in Sec.~\ref{sec:conclusion}.

\section{Preliminaries}
\label{sec:prelim}

\subsection{System Model}

Consider transmission  over the discrete time \ac{AWGN} channel
\begin{equation}
 Y_i = X_i + Z_i\label{eq:awgn}
\end{equation}
for $i=1,\ldots,n$. The channel input $X_i$ is taken from the $M$-ary
\ac{ASK} constellation $\cX$. The noise $Z_i$ is a Gaussian \ac{RV} with zero mean and
variance $\sigma^2$, i.e., $Z_i\sim\cN(0,\sigma^2)$. The \ac{SNR} is $1/\sigma^2$. All further results
are readily applicable to \ac{QAM} which has an \ac{ASK} constellation on the inphase and quadrature
component.

Mutual information is maximized under an average power constraint by a zero mean Gaussian input $X$ with unit variance,
and the capacity expression is
\begin{equation}
 \sfC_\tawgn(\text{SNR}) = \frac{1}{2}\log_2(1+\SNR)\label{eq:cap_awgn}.
\end{equation}

At the receiver, the decoder uses a decoding metric  $q^n(x^n, y^n): \cX^n \times \setR^n \to \setR^+$ to detect
the transmitted sequence $x^n$ in the codebook $\cC$ from the noisy observations $y^n$.
The decoding decision is $\hat x^n$ if
\begin{equation}
 \hat x^n = \argmax_{x^n \in \cC} q^n(x^n, y^n) = \prod_{i=1}^n q(x_i,y_i).
\end{equation}

In \cite{bocherer_achievable_2018}, it is shown that an achievable rate is
\begin{equation}
 R_\ta = \left[\entr(X) - \gentr(q)\right]^+\label{eq:ra}
\end{equation}
where $[\cdot]^+ = \max(0,\cdot)$  and $\entr(X)$ is the entropy of the discrete \ac{RV} $X$. The cross entropy $\gentr(\cdot)$ is the 
uncertainty a \ac{FEC} decoder needs to resolve\cite{bocherer_achievable_2018}:
\begin{equation}
 \gentr(q) = \E{-\log_2\left(\frac{q(X,Y)}{\sum_{x\in\cX} q(x, Y)}\right)}\label{eq:uncertainty}.
\end{equation}

For \ac{BMD}, we label each constellation point $x\in\cX$ with an $m$-bit label, 
i.e., $\chi : \cX \to \{0,1\}^m$ and $\chi(x) = b_1b_2\ldots b_m = \vb$. Its inverse is
$\chi^{-1}: \{0,1\}^m \to \cX$. A \ac{BRGC}~\cite{gray1953pulse} usually 
performs well for \ac{BMD} and the \ac{BMD} decoder uses the metric
\begin{equation}
 q(x,y) = q_\tbmd(\chi(x),y) = q_\tbmd(\vb,y) = \prod_{j=1}^m P_{B_j|Y}(b_j|y)\label{eq:bmd_metric}.
\end{equation}
Using~\eqref{eq:bmd_metric} in \eqref{eq:ra}, the achievable rate~\eqref{eq:ra} becomes
\begin{equation}
 R_\tbmd(\SNR; P_X) = \left[\entr(\vB) - \sum_{j=1}^m \entr(B_j|Y)\right]^+\label{eq:rbmd}.
\end{equation}
The \ac{FEC} decoder inputs are the soft information values
\begin{equation}
 l_j = \log\left(\frac{\sum_{x\in\cX_j^0} p_{Y|X}(y|x)P_X(x)}{\sum_{x\in\cX_j^1} p_{Y|X}(y|x)P_X(x)}\right)\label{eq:llr}, \quad j = 1, \ldots, m
\end{equation}
and $\cX_j^b = \{x\in\cX: [\chi(x)]_j = b\}$. 
For this choice, \eqref{eq:uncertainty} can be written as
\begin{equation}
 \gentr(q_\tbmd) = \sum_{j=1}^m\E{-\log_2\left(\frac{e^{(1-2B_j)(L_j/2)}}{e^{-L_j/2} + e^{L_j/2}}\right)}\label{eq:uncertainty_llr}.
\end{equation}

\subsection{Probabilistic Amplitude Shaping (PAS)}
\label{sec:PAS}

Probabilistic amplitude shaping (PAS)\acused{PAS} is a \ac{CM} scheme that combines \ac{PS} with \ac{FEC}~\cite{bocherer_bandwidth_2015}. It builds upon
two important properties. First, the capacity achieving distribution $P_X^*$ for the \ac{AWGN} channel is symmetric. We therefore
factor the input distribution into an amplitude and sign part as $P_X(x) = P_A(\abs{x}) \cdot P_S(\sign(x))$, where $P_A$ is non-uniform
on $\{\abs{x}, x\in\cX\}$ and $S$ is uniform on $\{-1,+1\}$. Second, the scheme exploits systematic encoding to preserve the non-uniform $P_A$.
It copies the binary representation of the amplitudes into the information part of the codeword and uses the approximately uniform
distributed parity bits as signs.

The \ac{DM}~\cite{schulte_constant_2016} realizes the non-uniform distribution $P_A$ on the amplitudes. It takes $k$ uniformly distributed
input bits and maps them to a length $n$ sequence of symbols with the empirical distribution $P_A$. For PAS, the \ac{DM} output set consists of amplitude values $\{\abs{x}, x\in\cX\}$.
The \ac{DM} rate is $R_\tdm = k/n$. The \ac{SE} of \ac{PAS} is~\cite[Sec.~IV-D]{bocherer_bandwidth_2015}
\begin{equation}
 \eta = R_\tdm + 1 - (1-R_\tc)\cdot m\label{eq:Rtx_PAS},
\end{equation}
for uniform signaling we have $R_\tdm = m-1$ and \eqref{eq:Rtx_PAS} becomes
\begin{equation}
 \eta = R_\tc \cdot m\label{eq:Rtx_uni}
\end{equation}
where $R_\tc$ is the code rate of the \ac{FEC} code.

\subsection{Low-Density Parity-Check (LDPC) Codes}
\label{sec:LDPC}

Binary \ac{LDPC} codes are linear block codes with a sparse parity-check matrix
$\vH \in \{0,1\}^{m_\tc\times n_\tc}$. \ac{LDPC} codes can be represented via their Tanner graph 
that are directly related to their parity-check matrices. A Tanner graph is a bipartite graph $G = (\cV \cup \cC, \cE)$
consisting of $n_\tc$ \acp{VN} and $m_\tc$ \acp{CN}. The set $\cE$ of edges contains the element
$e_{ij}$, denoting an edge between \ac{VN} $V_i\in\cV$ and \ac{CN} $C_j\in\cC$, if $h_{ji} = [\vH]_{ji} = 1$.
The sets $\cN(V_i)$ and $\cN(C_j)$ denote the neighbors of \ac{VN} $V_i$ and \ac{CN} $C_j$,
respectively. The degree $d_{\tv,i}$ ($d_{\tc,j}$) of \ac{VN} $V_i$ (\ac{CN} $C_j$) is 
the cardinality of the sets $\cN(V_i)$ and $\cN(C_j)$.

Practical \ac{LDPC} code designs are often based on \emph{protographs}~\cite{thorpe_protograph}.
The latter are defined via a basematrix $\vB$ of dimension $m_\tP \times n_\tP$ and $b_{ji} \in \setN$.
A basematrix may also be represented as a Tanner graph (also refered to as a protograph), however parallel edges
(corresponding to the multiplicity $b_{ji}$) are allowed. The final
parity-check matrix $\vH$ is obtained via a lifting or copy-and-permute operation, where a number of $n_\tc/n_\tP$ copies
of the original Tanner graph are placed next to each other and their edges are permuted such that connectivity constraints
imposed by the basematrix are maintained.

The proposed 802.3ca code has a basematrix with dimensions $12\times 69$ and an irregular degree profile of 
degree three, six, eleven and twelve \acp{VN}. The circulant size is 256, resulting in a final parity-check matrix
size of $m_\tc = 3072$ and $n_\tc = 17664$. The final graph has a girth of 6.
The degree 11 and 12 \acp{VN} of the protograph are punctured. While writing this manuscript,
the exact shortening pattern for the last information \ac{VN} is still being discussed. In the following, we assume this \ac{VN} to be shortened
completely. The number of transmitted bits is therefore $n_{\text{c,t}} = 16896$ with the overall code rate of $R_\tc = 14336/16896 \approx \num{0.8485}$. 

\subsection{Discretized Density Evolution (DDE)}
\label{sec:dde}

\ac{DDE} approximates real density evolution \cite{richardson_capacity} by discretizing the \ac{PDF} of the involved \ac{BP} messages. It was first used
to design capacity approaching \ac{LDPC} codes in \cite{chung_design} and quantizes the decoder soft-information \eqref{eq:llr} with a $b$ bit ($b\in\setN$) quantization
function, which first clips its input to $B \in \setR^+$ or $-B$ via $\textsf{clip}(\cdot)$ and maps the result to $q = 2^b-1$ quantization levels. We define this quantization function
as $\mathsf{Q}(\cdot): \setR \to \cQ$, where $\cQ = \{-(q-1)/2, \ldots, 0, \ldots, (q-1)/2\}$, $\Delta = (2B)/(q-1)$, and have
\begin{align}
 \mathsf{Q}(l) = \begin{cases} 
                  \left\lfloor \textsf{clip}(l)/\Delta + \frac{1}{2}\right\rfloor, & l > \frac{\Delta}{2}\\
                  \left\lceil \textsf{clip}(l)/\Delta - \frac{1}{2}\right\rceil, & l < -\frac{\Delta}{2}\\
                  0, & \text{otherwise}.
                 \end{cases}\label{eq:quant}                
\end{align}
We use this type of quantization to represent $l = 0$ without quantization error. This is important for punctured \acp{VN}.

\section{Optimizing the Bit Mapping for Off-the-Shelf Protograph LDPC Codes}
\label{sec:optim_mapping}

\subsection{Problem Formulation}
\label{sec:optim_mapping_problem_formulation}

The bit channels $p_{L_j|B_j}$ resulting from \ac{BMD} have different qualities. Previous works noted the importance of 
mapping the different \ac{BMD} bit levels to the different \ac{VN} degrees of an irregular \ac{LDPC} code, or to \ac{VN} types of a protograph
to improve the performance~\cite{divsalar_protograph_2005-3,richter_mapping_2007,jin_optimized_2010,cheng_exit-aided_2012,hager2014optimized,hager_improving_2014}.

In the following, we use the ideas of \cite{cheng_exit-aided_2012,hager2014optimized} to formulate an optimization procedure 
that optimizes the assignment of the $m$ bit channels to the $n_{\tP,t}$ transmitted \ac{VN} types of a given protograph basematrix.
This bit mapping can be expressed as a non-negative matrix $\vA = [\va_1, \ldots, \va_{n_{\tP,t}}]$ of dimension $m \times n_{\tP,t}$ where the entry $a_{ji} = [\vA]_{ji}$ denotes
the fraction of bit level $j$ that is assigned to the $i$-th transmitted \ac{VN} type. The matrix $\vA$ needs to fulfill
the constraints
\begin{align}
 \sum_{i'=1}^{n_{\tP,t}} a_{ji'} \frac{1}{n_{\tP,t}} &= \frac{1}{m}, & \sum_{j'=1}^{m} a_{j'i} &= 1,
\end{align}
for all $j \in \{1,2,\ldots, m\}$ and $i \in \{1,2,\ldots, n_{\tP,t}\}$. We denote the set of matrices $\vA$ which fulfill 
the above constraints as $\cA$. For PAS, we further impose the constraints
\begin{align}
 a_{1i} &= 1, &  a_{ji} &= 0, \quad j \in \{2,\ldots, m\}, \quad i \in \cP\label{eq:A_constraints_PAS}
\end{align}
where $\cP$ is the set of transmitted parity \acp{VN} in the protograph basematrix, as the
parity bits have to be mapped to bit level one. The set $\cA$ is adjusted accordingly in this case.

Let the \ac{BP} decoding threshold for a given basematrix $\vB$, bit mapping $\vA$ and signaling mode $P_X$ be $\SNR^*(\vA; \vB, P_X)$. The optimization problem is
\begin{align}
 \min_{\vA}\quad \SNR^*(\vA; \vB, P_X) \qquad \text{subject to } \vA \in \cA\label{eq:full_optim}.
\end{align}
The optimization requires an efficient way to evaluate the objective for a given parameter set. The obvious choice is \ac{DDE}; 
however, calculating the decoding threshold via \ac{DDE}
takes a couple of seconds for the considered protograph sizes. Its use as part of an optimization algorithm which evaluates the objective
many times is therefore limited. Instead, we combine the approaches of \cite{hager2014optimized,steiner_protograph-based_2016}
and use P-EXIT~\cite{liva_protograph_2007}. In \cite{steiner_protograph-based_2016}, the $m$ \ac{BMD} bit channels are matched to $m$ parallel
binary-input AWGN (biAWGN)\acused{biAWGN} surrogate channels for which a P-EXIT analysis is feasible. To perform this matching, we use the conditional entropy $\entr(B_j|Y)$, and the
corresponding \ac{biAWGN} surrogate channel $\tilde Y_j = \tilde X_j + \tilde Z_j$, with input $\tilde X_j\in\{-1,+1\}$ and noise $\tilde Z_j \sim \cN(0,\tilde\sigma_j^2)$ that is determined
by solving
\begin{align}
 \tilde\sigma_j^2: \entr(\tilde X_j|\tilde Y_j) = \entr(B_j | Y), \quad j = 1, \ldots, m\label{eq:pexit_matching}. 
\end{align}
For a given mixing vector $\va = (a_1, \ldots, a_m)$, i.e., a column of $\vA$, we find the \ac{biAWGN} surrogate with parameter $\tilde\sigma^2_j$ as
\begin{align}
 \tilde\sigma_j^2: \entr(\tilde X_j|\tilde Y_j) = \sum_{j'=1}^m a_{j'} \entr(B_{j'} | Y), \quad j = 1, \ldots, m\label{eq:pexit_matching_general}. 
\end{align}

\subsection{Successive Bit Mapping Optimization}
\label{sec:successive_optim}

Performing the optimization \eqref{eq:full_optim} jointly over all bit levels is a complicated task, as it involves a large number of optimization variables
for large constellation sizes and protograph \acp{VN}. Instead, we propose a successive method that optimizes each bit level one at a
time while leaving the mappings of the other bit levels fixed. As a consequence, we do not optimize over the whole bit mapping matrix $\vA$, but only over one row of $\vA$,
where the ordering is chosen as a parameter. All other bit levels are assigned uniformly. The algorithm for uniform signaling is summarized in Algorithm~\ref{alg:successive_optim}. For PAS,
the function \textsc{make\_A} is modified accordingly to account for the additional constraints~\eqref{eq:A_constraints_PAS}.
For the optimization in line 3, we use \emph{patternsearch}~\cite{hooke_direct_search_1961}, a derivative free optimization approach that starts from a feasible initial point $\vx$
(i.e., one that fulfills the constraints) and then performs a search with a set of vectors to find a direction in which the objective value improves.
For our setting, we use a so-called $2N$ basis which consists of the $2N$ canonical unit vectors $\ve_i, i = 1, \ldots, N$ of $\setR^N$ and their negative counterparts,
where $N$ is the number of independent optimization variables. The algorithm then polls all possible new points $\vx \pm s \cdot \ve_i$ after an appropriate scaling ($s\in\setR^+$) of the basis vectors
and selects the one with the best objective value as the starting point for the next iteration.

\begin{algorithm}[t]
\footnotesize
\centering
\begin{algorithmic}[1]
\INPUT Protograph $\vB$, Distribution $P_X$, Ordering $\cO$, Set of fixed row indices $\cI$
\State $\vA_{\tF} \gets [], \cI \gets []$
\For{$j \in \cO$}
    \State $\va_{\topt} = \argmin_{\va} \SNR^*(\textsc{make\_A}(\va, j, \vA_\tF, \cI); \vB, P_X)$ subject to $0 \leq a_i \leq 1 - \text{sum}(\vA_\tF(:,i),1)), \forall i \in \{1, \ldots, n_{\tP,t}\}$
    \State $\vA_{\tF} \gets \vectb{\vA_{\tF}\\ \va_{\topt}}$, $\cI \gets \cI \cup \{j\}$
\EndFor
\State $\vA_\topt = \textsc{make\_A}(\{\}, \{\}, \vA_{\tF}, \cI)$

\Function{make\_A}{$\va, j, \vA_\tF, \cI$}
\State $\vA(j,:) \gets \va$
\State $\vA(\cI,:) \gets \vA_\tF$
\State $\vA([1:m]\setminus\cI,:) = (1/(m-\abs{\cI})) \cdot (1 - \text{sum}(\vA_\tF,1))$
\State \Return $\vA$
\EndFunction
\end{algorithmic}
\caption{Algorithmic description of the successive bit mapping optimization.}
\label{alg:successive_optim}
\end{algorithm}

\subsection{Numerical Results}
\label{sec:bit_mapper_optim_results}

We focus on a scenario with an \ac{SE} of $\eta = \num{2.545}$ bits per channel use (bpcu). Uniform signaling uses 8-ASK, whereas PAS uses 16-ASK with an 
appropriately chosen \ac{MB} input distribution~\cite{kschischang_pasupathy_maxwell}. The different constellation sizes are chosen such that the best performance for both signaling modes is ensured. The \ac{DM} rate for PAS is $R_\tdm = \SI{2.152}{bits}$.

We validate the P-EXIT thresholds by \ac{DDE} and use a 8-bit quantization ($q=255$) with $B=15$. These values are motivated by the observations in Sec.~\ref{sec:opt_clipping},
which show that a decoder with these parameters operates with almost no loss as compared to a full resolution, floating point implementation. The obtained decoding thresholds are 
summarized in Table~\ref{tab:cmp_pexit_dde}. As a reference we choose a bit mapping $\vA_\text{ref}$ which assigns each bit level uniformly
to each \ac{VN} type, i.e., $\vA_\text{ref} = 1/m \cdot \bm{1}$, where $\bm{1}$ is the all-ones matrix of size $m\times n_{\tP,t}$. For PAS, $\vA_\text{ref}$ is additionally adjusted to meet the constraints of \eqref{eq:A_constraints_PAS}.

We observe that the P-EXIT and \ac{DDE} values are in
good agreement with a maximum difference of \SI{0.12}{dB}, which is caused by the surrogate analysis and the mixing of the bit channels.
\begin{table}
\centering
\footnotesize
 \caption{Comparison of decoding thresholds with P-EXIT and \ac{DDE} for PAS and uniform signaling.}
 \begin{tabular}{llll}
  \toprule
  Signaling & Bit mapping  & P-EXIT [\si{dB}] & DDE [\si{dB}]\\
  \midrule
  \multirow{2}{*}{PAS} & reference & \num{16.00} & \num{16.12}\\
                      & optimized & \num{15.67} & \num{15.75}\\
  \multirow{2}{*}{uniform} & reference & \num{17.12} & \num{17.21}\\
                      & optimized & \num{16.90} & \num{16.98} \\
  \bottomrule
 \end{tabular}
 \label{tab:cmp_pexit_dde}
\end{table}
For uniform signaling the gain is \SI{0.23}{dB}, and for PAS the gain is \SI{0.37}{dB} based on the \ac{DDE} thresholds. In both cases, the optimization yields a bit mapping $\vA_\topt$, 
which favors the assignment of the most reliable bit-channel (i.e., the one with the smallest uncertainty $\entr(B_j|Y)$) to the degree six \acp{VN} in the protograph.
For uniform signaling, this means bit-level one is mapped to the degree 6 \ac{VN} types, whereas for PAS bit level two (which has the largest prior $\log(P_{B_j}(0)/P_{B_j}(1))$) is the
most reliable one and is assigned to them. Empirical studies show that the ordering $\cO$ (cf. the input of Algorithm~\ref{alg:successive_optim}) plays an important role and 
that the best decoding threshold is achieved by starting with the bit channel having the smallest uncertainty. This result is intuitive as the first bit channel
has the largest degree of freedom for the bit mapping optimization. We validate the asymptotic results by finite length simulations in Fig.~\ref{fig:cmp_uni_shaped}.

\begin{figure*}
 \centering
 \footnotesize
 \subfloat[][$q = 7$ (3-bit quantization)]{\begin{tikzpicture}
\begin{axis}[
xlabel={$B$},
ylabel={Required SNR [\si{dB}]},
grid=both,
legend cell align=left,
xtick={4, 5, ..., 16},
ytick={15, 15.5, ..., 19},
ymin=15,
ymax=19.5
]


%

\addplot[line width=1,black] table[x=B,y=snr_req] {data/opt-B-rates-q=7.txt};\label{plt:opt_B_rates}


\addplot[line width=1,TUMBeamerBlue,dashed] table[x=B,y=snr_req] {data/opt-B-th-uni-consecutive-q=7.txt};\label{plt:opt_B_uni_th_consecutive}
\addplot[line width=1,TUMBeamerBlue] table[x=B,y=snr_req] {data/opt-B-th-uni-opt-q=7.txt};\label{plt:opt_B_uni_th_opt}

\addplot[line width=1,TUMBeamerBlue,mark=o,dashed,mark options={solid}] table[x=B,y=snr_req] {data/opt-B-sim-uni-consecutive-q=7.txt};\label{plt:opt_B_uni_sim_consecutive}
\addplot[line width=1,TUMBeamerBlue,mark=o] table[x=B,y=snr_req] {data/opt-B-sim-uni-opt-q=7.txt};\label{plt:opt_B_uni_sim_opt}


\addplot[line width=1,TUMBeamerRed,dashed] table[x=B,y=snr_req] {data/opt-B-th-shaped-consecutive-q=7.txt};\label{plt:opt_B_shaped_th_consecutive}
\addplot[line width=1,TUMBeamerRed] table[x=B,y=snr_req] {data/opt-B-th-shaped-opt-q=7.txt};\label{plt:opt_B_shaped_th_opt}

\addplot[line width=1,TUMBeamerRed,mark=o,dashed,mark options={solid}] table[x=B,y=snr_req] {data/opt-B-sim-shaped-consecutive-q=7.txt};\label{plt:opt_B_shaped_sim_consecutive}
\addplot[line width=1,TUMBeamerRed,mark=o] table[x=B,y=snr_req] {data/opt-B-sim-shaped-opt-q=7.txt};\label{plt:opt_B_shaped_sim_opt}


\draw[black,line width=1,dashed] (axis cs: 2, 15.25) -- (axis cs: 20, 15.25);



\node at (axis cs: 14.5, 15.45) {$R_{\tbmd}^{-1}(\eta;P_X)$};

\end{axis}
\end{tikzpicture}}
 \subfloat[][$q = 15$ (4-bit quantization)]{\begin{tikzpicture}
\begin{axis}[
xlabel={$B$},
ylabel={Required SNR [\si{dB}]},
grid=both,
legend cell align=left,
xtick={4, 5, ..., 16},
ytick={15, 15.5, ..., 19},
ymin=15,
ymax=19.5
]


\addplot[line width=1,black] table[x=B,y=snr_req] {data/opt-B-rates-q=15.txt};
%


\addplot[line width=1,TUMBeamerBlue,dashed] table[x=B,y=snr_req] {data/opt-B-th-uni-consecutive-q=15.txt};
\addplot[line width=1,TUMBeamerBlue] table[x=B,y=snr_req] {data/opt-B-th-uni-opt-q=15.txt};

\addplot[line width=1,TUMBeamerBlue,mark=o,dashed,mark options={solid}] table[x=B,y=snr_req] {data/opt-B-sim-uni-consecutive-q=15.txt};
\addplot[line width=1,TUMBeamerBlue,mark=o] table[x=B,y=snr_req] {data/opt-B-sim-uni-opt-q=15.txt};


\addplot[line width=1,TUMBeamerRed,dashed] table[x=B,y=snr_req] {data/opt-B-th-shaped-consecutive-q=15.txt};
\addplot[line width=1,TUMBeamerRed] table[x=B,y=snr_req] {data/opt-B-th-shaped-opt-q=15.txt};

\addplot[line width=1,TUMBeamerRed,mark=o,dashed,mark options={solid}] table[x=B,y=snr_req] {data/opt-B-sim-shaped-consecutive-q=15.txt};
\addplot[line width=1,TUMBeamerRed,mark=o,] table[x=B,y=snr_req] {data/opt-B-sim-shaped-opt-q=15.txt};


\draw[black,line width=1,dashed] (axis cs: 2, 15.25) -- (axis cs: 20, 15.25);



\node at (axis cs: 14.5, 15.45) {$R_{\tbmd}^{-1}(\eta;P_X)$};

\end{axis}
\end{tikzpicture}}
 \caption{Optimal value of $B$ based on the uncertainty, \ac{DDE} decoding thresholds and finite length simulation results. The target \ac{SE} is $\eta = \SI{2.545}{\bpcu}$ and we depict the required \SNR{} to achieve this \ac{SE} for \ac{PAS} and uniform signaling. 
 The black curves (\ref{plt:opt_B_rates}) are based on the uncertainty \eqref{eq:uncertainty_llr}. The curves without markers denote the \ac{DDE} decoding thresholds for uniform (reference mapping: \ref{plt:opt_B_uni_th_consecutive}, optimized
 mapping: \ref{plt:opt_B_uni_th_opt}) and PAS signaling (reference: \ref{plt:opt_B_shaped_th_consecutive}, optimized: \ref{plt:opt_B_shaped_th_opt}). The curves with markers are the corresponding finite length simulation results for a frame error rate of
 \num{e-3}.}
 \label{fig:optim_B}
\end{figure*}
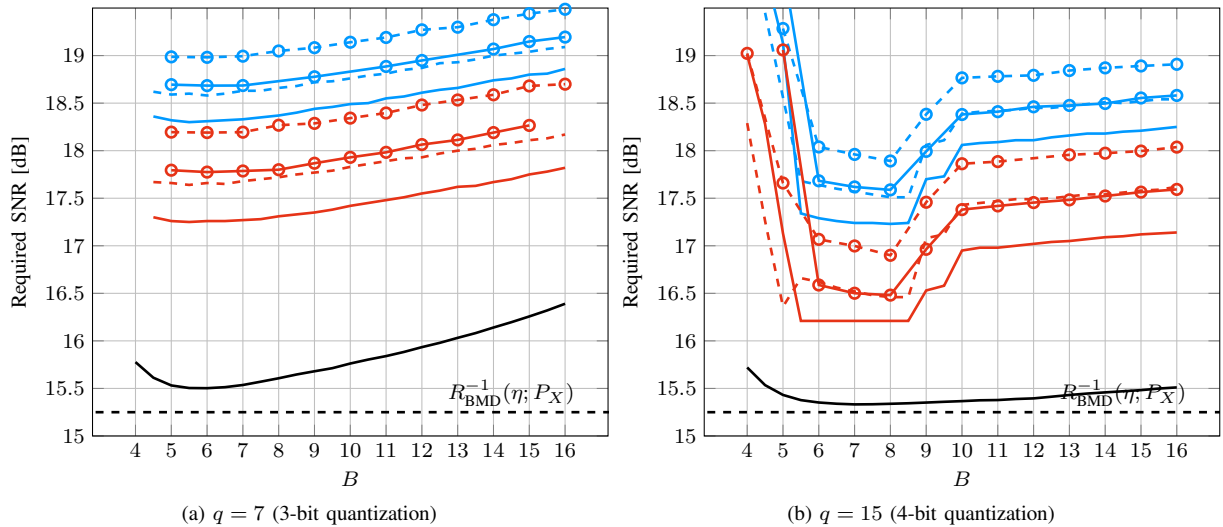

\section{Clipping Optimization for Quantized Decoders}
\label{sec:opt_clipping}

We now investigate the influence of the number of quantization levels $q$ and the clipping $B$. As noted in \cite{zhao_implementation_2005}, clipping the
soft-information can greatly influence the performance of the decoder and depends on the considered code
ensemble.

We examine two scenarios with $b=3$ and $b=4$ bits resolution and investigate different
approaches to find the best $B$. The first approach considers the uncertainty expression in \eqref{eq:uncertainty_llr}.
We evaluate the metric by generating soft-information values according to \eqref{eq:llr}, quantizing them~\eqref{eq:quant}
and approximating the expectation by its empirical mean in a Monte-Carlo manner. The second approach uses \ac{DDE},
as defined in Sec.~\ref{sec:dde}, and determines the decoding threshold of the \ac{LDPC} code ensemble given the selected
quantization and clipping parameters.

We depict the results of this analysis in Fig.~\ref{fig:optim_B} for the setting of Sec.~\ref{sec:bit_mapper_optim_results}. The
optimized clipping is given by $B \approx 6$ for three bits and by $B \approx 8$ for four bits. The lines without markers represent the \ac{DDE} thresholds,
whereas the lines with markers are finite length simulation results and denote the required \ac{SNR} to obtain a target FER of
\num{e-3}. Observe that the simulation results closely follow the \ac{DDE} thresholds. Observe also that the uncertainty 
provides a good indication for the optimal clipping value, but does not reflect the overall qualitative behavior. 

In Fig.~\ref{fig:cmp_uni_shaped}, we show the \ac{BER} curves for the quantized decoders discussed in this section.
We see that the loss due to quantization is about \SI{0.25}{dB} for 4 bits and \SI{1.50}{dB} for 3 bits. A quantized decoder
with $B=15$ and $q=8$ operates with almost no loss as compared to a floating point implementation with full double resolution.

\section{Conclusion}
\label{sec:conclusion}

Using the example of the proposed IEEE 802.3ca \ac{LDPC} code, we have shown how off-the-shelf protograph based \ac{LDPC} codes can be optimized for higher-order modulation
and a quantized decoder. The optimized bit mapping improves the finite-length performance
by \SI{0.4}{dB} for PS and by \SI{0.3}{dB} for uniform signaling without increasing the complexity. We also found the best clipping values for a quantized decoder 
with \ac{DDE} and showed its accuracy with finite length simulations. Future work can further optimize the quantization levels and \ac{LUT} entries for the
\ac{CN} operation.

\vspace{-.45\baselineskip}

\end{document}